\documentclass{eas}
\usepackage{graphicx}
\usepackage{astron}
\begin{document}

\title{An N band interferometric survey of the disks around post-AGB binary stars} 
\runningtitle{A MIDI survey of post-AGB disks}
\author{Michel Hillen}\address{Institute of Astronomy, University of Leuven; \email{Michel.Hillen@ster.kuleuven.be}}

\begin{abstract}
It is now well established that FGK post-AGB stars that are surrounded by both hot and cold dust (as derived from the
spectral energy distribution), are almost always part of a binary system with $100 < P_{orb} < 5000$~days.
The properties and long-term stability 
of the dust emission requires it to arise from a gas- and dust-rich, puffed-up and (semi-)stable circumbinary disk. 
This interpretation has been confirmed with spatially resolved observations at a range of wavelengths for various individual objects. 
Here I present the first results of the first mid-IR interferometric survey of this class of objects. 
Our sample comprises 18 sources, most of which are confirmed binaries and which cover a range in IR excess. 
Our analysis clearly shows the compactness of the dust structures in 
these systems. We perform a statistical comparison with radiative transfer disk models, showing that most objects are indeed
continuous disks from the sublimation radius outwards.
\end{abstract}
\maketitle
\section{Introduction}
To commemorate Olivier Chesneau I would like to start this paper with a very relevant quote expressed by him in Chesneau \cite*{2013LNPChesneau}:

\begin{quotation}
``My personal opinion is that the discovery of a stratified disk with proved Keplerian kinematics is 
directly connected to the influence of a companion, albeit the few exceptions \ldots This hypothesis must be confirmed by
further observations.''
\end{quotation}

Here I want to highlight some of the observational evidence that has been collected 
in recent years to confirm this hypothesis in the particular case of post-AGB stars\footnote{Stellar remnants, with spectral types of (A)FGK(M), of 
low- to intermediate mass stars.}. We have come to a point where we can state that ``it 
is observationally established that the formation of a Keplerian stratified disk around an object in the post-AGB 
evolutionary phase \textit{requires} the influence of a close companion star''. The more interesting questions now become, 1) how 
do these disks form and evolve, and affect the central stars, and 2) can these disks teach us something about 
the evolution of dusty disks in general, about the mechanism behind angular momentum transport, about the conditions 
needed for planet formation, \ldots ?

\section{The disk-binary connection}
\begin{figure}
   \centering
   \includegraphics[width=9cm]{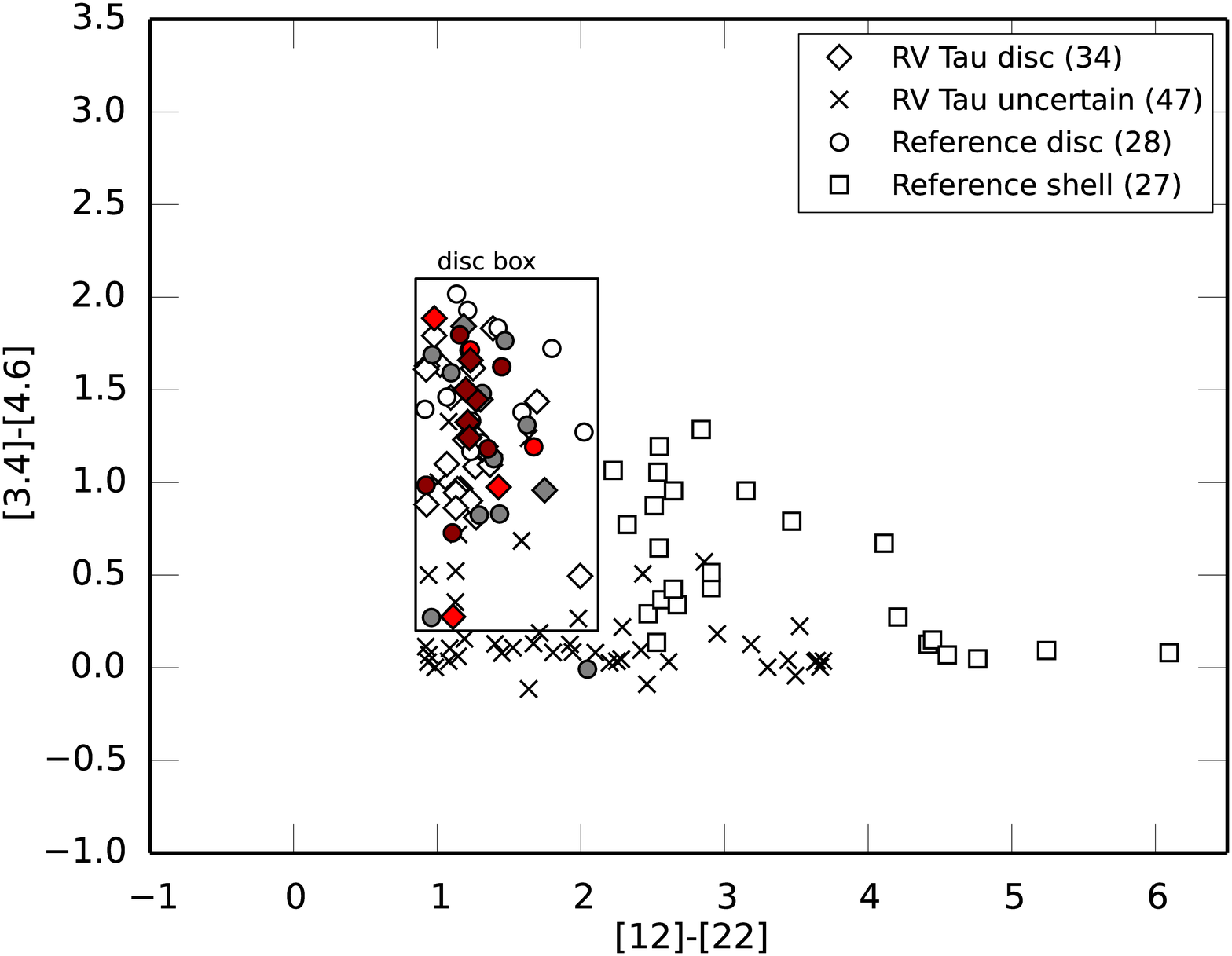}
   \caption{A WISE color-color diagram of a large sample of post-AGB stars, including all the RV Tauri stars in the GCVS with an infrared excess
   (based on the work of Gezer et al., 2015). The circles and squares are the reference samples of disks and shells, respectively. Diamonds 
   are RV Tauri stars that can be reliably classified from the SED as disks, based on their similarity with the reference sample objects. 
   Crosses are objects with an uncertain classification for various reasons.
   Grey and dark-red symbols refer to the confirmed (spectroscopic) binaries. Dark- and bright-red symbols are the sources in 
   the N-band interferometric survey with reliable WISE fluxes.}
   \label{figure:sample}
\end{figure}

One of the simplest arguments for the disk interpretation of the infrared excess around a large fraction of the known 
post-AGB stars \cite{2014MNRASKamath} is that they are too hot and too evolved to still produce and expell dust \cite{2003ARAAVanWinckel}. 
The shape of the infrared excess is a critical quantity to distinguish between a post-AGB star that is surrounded by an expanding 
shell or by a Keplerian disk: the latter have a characteristic near-IR excess because some of the dust is orbiting close to the 
central object. The difference in the shape of the excess is well summarized by the WISE color-color 
diagram in Fig.~\ref{figure:sample}, which was constructed after the work of Gezer {\em et al.\/}~\cite*{2015arXivGezer}.
The diagram includes a large fraction of the Galactic optically-bright post-AGB stars, including all the RV Tauri stars in the General 
Catalogue of Variable stars with an infrared excess. RV Tauri stars are Type II Cepheids with pulsation periods above 20 days, likely in a 
post-AGB evolutionary phase \cite{2002PASPWallerstein}. Circles and squares are objects that were reliably classified in the literature 
as either disks or shells, respectively. The region occupied by disk sources can be arbitrarily referred to as the ``disk box''. Many RV Tauri
stars also have IR colors that fall within the disk box, with a distribution that is indistinguishable from the reference sample. 
Whether a post-AGB star pulsates clearly does not have an influence on whether a disk may be present around it.

Gezer {\em et al.\/}~\cite*{2015arXivGezer} also correlated the excess with other properties of the central object, like the photospheric 
abundance and the question of duplicity. As Fig.~\ref{figure:sample} shows, there is a clear correlation with the latter: all confirmed binaries 
are in the disk box. There may be an observational bias involved, but some shell sources have been extensively monitored for duplicity without
success \cite{2011ApJHrivnak}. Most objects within the disk box are being monitored with the Mercator telescope and 
the detected binary fraction in this region of the color-color diagram will increase in the near future (Manick {\em et al.\/}, in prep.).

Confirmation of the disk nature of objects in the ``disk box'' requires multiwavelength high-angular-resolution observations. 
For example, the Keplerian kinematics has now been resolved in CO rotational lines with sub-mm interferometry for two objects 
\cite{2015AABujarrabal,2013AABujarrabalC}. 

\section{Results of the N band interferometric survey}
Spatially resolved data also allow one to look for signatures of disk evolution. The mid-IR spectral window is well-suited 
for this, since the infrared excess of a continuous disk starting at the sublimation radius peaks at these wavelengths. 
Disks with gaps or large inner holes have depressed near- to mid-IR fluxes (as is the case for AC Her, Hillen {\em et al.\/}~\cite*{2015AAHillen}).

\begin{figure}
   \centering
   \includegraphics[width=9cm]{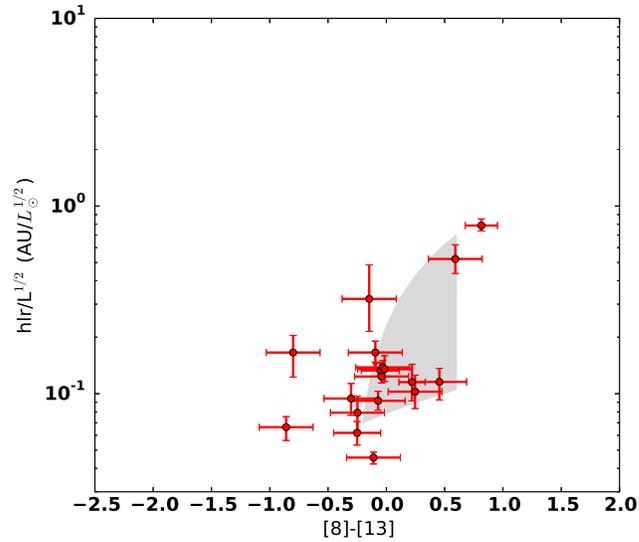}
   \caption{The distance-independent color-size diagram derived from the MIDI data. Horizontally the color 
   of the IR excess over the N band is shown; vertically the half-light-radius of the 10~$\mu m$ emission 
   normalised with the square root of the stellar luminosity.
   In grey is the region occupied by the radiative transfer models of Menu {\em et al.\/}~(2015).}
   \label{figure:color-size}
\end{figure}

We collected all the VLTI/MIDI data of post-AGB stars in the disk box (a total of 18 sources, 14 of which are included 
in Fig.~\ref{figure:sample}). Each source typically has a few measurements, just enough to derive the typical size of the mid-IR emission.
We follow the same modeling approach as Menu {\em et al.\/}~\cite*{2015arXivMenu}, the details of which will be described in 
a dedicated paper (Hillen et al., in prep.). Fig.~\ref{figure:color-size} shows a sneak preview of the main result of the survey, which is a 
mid-IR color-size diagram. In addition to the sample sources, we also include the region 
occupied by radiative transfer continuous disk models, computed by Menu {\em et al.\/}~\cite*{2015arXivMenu} 
to represent their Herbig Ae protoplanetary disks. Most objects cluster near the region delimited by the models, underlining the 
disk nature of our sources. A more proper analysis with dedicated post-AGB disk models will be conducted to draw definitive conclusions
from this rich interferometric data set.

\bibliography{definitions,Hillen_NbandSurvey}
\bibliographystyle{astron}

\end{document}